\documentclass[
    ,final            
  ]
  {aipproc}

\layoutstyle{6x9}

\begin{document}

\title{Analyzing Powers for Forward $p_\uparrow+p\rightarrow\pi^0+X$ 
            at STAR
\footnote{Talk delivered at the ``15th International Spin Physics
 Symposium,'' SPIN2002, September 9-14, 2002, Brookhaven National
 Laboratory, Upton, NY, USA.}}

\author{G.\ Rakness for the STAR Collaboration}{
  address={Indiana University Cyclotron Facility \\
           Bloomington, IN  47408 USA}
}

\begin{abstract}
Preliminary results of the analyzing power for the production of forward, 
high-energy $\pi^0$ mesons from collisions of transversely polarized 
protons at $\sqrt{s}=200\,$GeV from STAR are presented.
The kinematic ranges covered by the data are $x_F\approx 0.2-0.6$ and
$p_T\approx 1-3\,$GeV/c.
The analyzing power at $\sqrt{s}=200\,$GeV is found to be comparable 
to that observed at $\sqrt{s}=20\,$GeV.

\end{abstract}

\maketitle


\section{Introduction}

Perturbative QCD makes a qualitative prediction that
the single-spin transverse asymmetry, known as the analyzing power 
($A_N$), for $2\rightarrow 2$ parton scattering at large transverse 
momentum should be zero based on helicity conservation.
In the late 1980's, the experiment E704 at Fermi National Laboratory 
measured $A_N$ for the production of 
charged and neutral pions in $p_\uparrow+p$ collisions
at $\sqrt{s}=20\,$GeV and $p_T=1-3\,$GeV/c
over the Feynman-$x$ range of $0-0.8$~\cite{E7041,E7042}.
The analyzing
power for $\pi^+$ mesons was found to increase as a function
of Feynman-$x$ from $A_N=0$ at $x_F=0$ to $A_N\approx 0.4$ at $x_F=0.8$.
For $\pi^-$ mesons, the analyzing power was found to be
approximately equal in magnitude and opposite in sign to the $\pi^+$ 
results, while for $\pi^0$ mesons, the analyzing power was found to be
approximately half the size observed for $\pi^+$ mesons.

This result has inspired several theory groups to develop
models to account for the observed large analyzing power.
Most models attribute forward pion production to collisions between 
a quark in one proton and a gluon in the other.
There are many  
different plausible mechanisms by which one might expect transverse spin
effects, all of which could contribute to some degree.
One perturbative QCD approach attributes
transverse spin effects to twist-3 gluon correlations 
before or after the primary
quark-gluon coupling~\cite{qiusterman,koike}.
A second approach attributes $A_N$ to the transversity 
distribution function and a T-odd Heppelmann-Collins
fragmentation function~\cite{anselminocollins}.
Another approach is to include initial state interactions to introduce
transverse spin effects before the primary quark-gluon 
coupling~\cite{anselminosivers1,anselminosivers2}.
All of these models predict the large analyzing power observed at 
E704 should persist to collision
energies an order of magnitude greater~\cite{review,dalesio}.
Here we present the measurement of the analyzing power 
for the production
of $\pi^0$ mesons from the STAR experiment at the Relativistic Heavy
Ion Collider (RHIC) at Brookhaven National Laboratory, studying 
$p_\uparrow + p$ collisions with total energy
$\sqrt{s}=200\,$GeV available to the system.

\section{Experimental Conditions}

The analyzing power for a reaction with a transversely polarized beam 
interacting with an unpolarized target is determined from a 
spin-dependent asymmetry,
\begin{equation}
  \cos\phi\, P_{beam}\, A_N = \frac{N_+-RN_-}{N_++RN_-},
\label{asym}
\end{equation}
and requires the concurrent measurements of three independent 
quantities.
The magnitude of the transverse polarization of the beam is $P_{beam}$.
The number of measured $\pi^0$ events observed when the polarization 
direction is up(down) is $N_{+(-)}$.
The relative luminosity is given by 
$R={\mathcal L_+}LT_+/{\mathcal L_-}LT_-$, where $\mathcal{L_{+(-)}}$ 
is the luminosity and $LT_{+(-)}$ is the livetime of the detector 
for different polarization states.
The spin-dependent asymmetry corresponds to the right-hand side of
equation~\ref{asym}.
The azimuthal angle between the polarization vector and the normal
to the reaction plane is $\phi$.
Parity constrains the asymmetry to be zero when the $\pi^0$ is emitted 
along the direction of the polarization vector.

Data were collected during the polarized proton run at RHIC in January 
2002, and, as such, resulted from the first observations of polarized 
protons in a collider environment.
A typical RHIC fill lasted $6-8\,$hours with collision luminosities on 
the order of $10^{30}\,$cm$^{-2}$sec$^{-1}$ at the STAR interaction 
region.
The so-called ``yellow'' proton beam rotated counterclockwise around RHIC 
when viewed from above, while the ``blue'' proton beam went clockwise.
Each beam contained 55 filled bunches and 5 empty bunches
which collided every $213\,$nsec at the STAR interaction region,
resulting in 50 chances for collisions per revolution.
The beam polarization vector was oriented vertically, perpendicular to
the proton momentum direction.
In the polarization pattern for the yellow beam, the polarization direction
alternated every bunch, while the blue pattern alternated every second
bunch.
The data presented here refer to spin asymmetries with respect to the
direction of the yellow polarization vector, averaging over the blue 
polarization.  

The average beam polarization for each fill was given by the 
Coulomb-Nuclear Interference (CNI) polarimeter located at 
12 o'clock in RHIC~\cite{Kurita,Jinnouchi}.
This detector measured the asymmetry of elastic proton-carbon 
collisions by observing recoil carbon atoms at approximately 90 degrees.
At the RHIC injection energy ($\approx 25\,$GeV per beam), the 
analyzing power of the CNI reaction has been measured~\cite{E925} 
and can be used to deduce the absolute polarization of the proton 
beam.
At RHIC collision energies ($\approx 100\,$GeV per beam),
the analyzing power for this reaction has not yet been measured.
Therefore, for these proceedings the beam polarization is determined
by using the average value of the measured CNI 
asymmetry at collision energy within each fill, divided by 
the measured analyzing power of the CNI reaction at injection energy.
With this assumption, the average luminosity-weighted beam polarization 
for the data presented here is $P_{beam}\approx 16\%$.
As the asymmetry measured at collision energy was consistent with
the asymmetry measured at injection energy for these runs, the 
$P_{beam}$ quoted here represents a likely upper limit at the 
collision energy, since the beam acceleration is not expected to 
enhance $P_{beam}$.

\begin{figure}
  \includegraphics[height=3.0in]{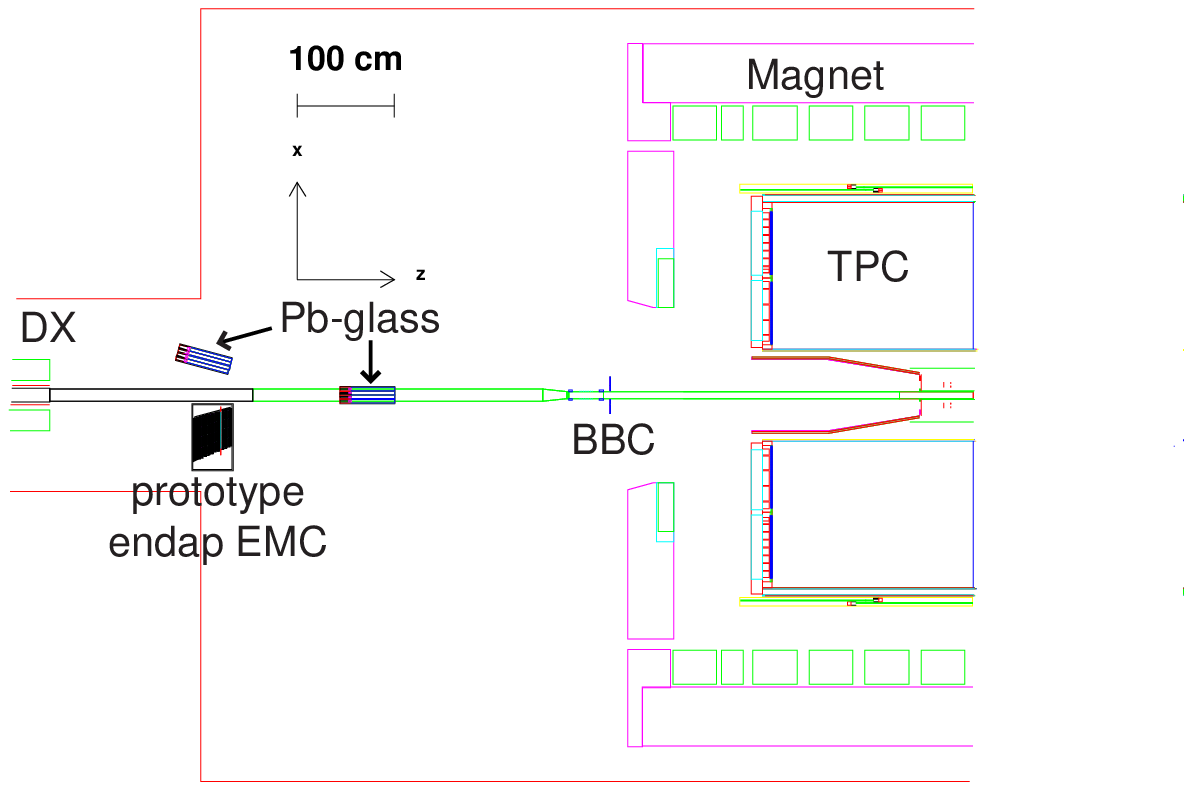}
  \caption{Schematic of the FPD and the east half STAR, looking from above.
           In this figure, the yellow beam moves from right to left, and 
           collisions occur at the right-hand edge.  
           }
  \label{schematic}
\end{figure}

A forward $\pi^0$ detector (FPD) comprising four arms was installed 
at STAR approximately $750\,$cm from the interaction region and very 
close to the beam pipe (Fig.~\ref{schematic}).
Its location was such that significant energy deposition corresponded to 
positive $x_F$ particle production with respect to the yellow beam.
An electromagnetic Pb-scintillator sampling calorimeter of $\approx 21$ 
radiation lengths subdivided into 12 towers was placed to the left of 
the oncoming yellow beam.
This detector is a prototype of 1/60 of the endcap 
electromagnetic calorimeter (pEEMC), currently being installed at STAR.
The pEEMC has two layers of preshower readout and a shower-maximum detector 
(SMD) made of orthogonal layers of finely segmented scintillator strips to 
measure the longitudinal and transverse profiles of photon showers.
A $4\times 4$ array of $3.8\times 3.8\times 45\,$cm$^3$ Pb-glass detectors 
was placed to the right of the oncoming beam as well as above and below 
the beam.
Readout of all FPD calorimeters was triggered for events that deposited
$\approx 20\,$GeV electron-equivalent energy in any one calorimeter.
The kinematic ranges covered by the FPD were $1<p_T<3\,$GeV/c and 
$0.2<x_F<0.6$.
A valid coincidence from scintillator annuli mounted around the beam on 
both sides of the STAR magnet was required in the offline analysis of 
the data.
These scintillator annuli are called the STAR beam-beam counters 
(BBC)~\cite{joanna}.
A dedicated beam study in which the beams were steered out of collision 
at the STAR interaction region determined that approximately 98\% of the 
observed FPD triggers accompanied by a BBC coincidence came as a result 
of $p+p$ collisions.

\section{Data Analysis}

\begin{figure}
  \includegraphics[height=3.0in]{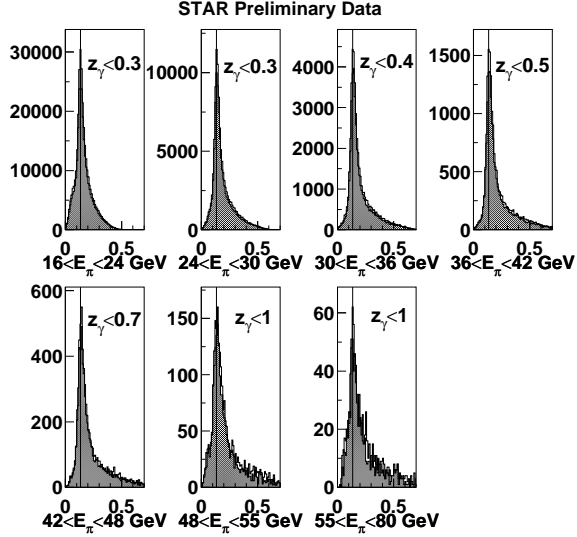}
  \caption{Preliminary distributions of the diphoton invariant mass 
    spectra sorted into the energy bins 
    used in the asymmetry analysis.
    The distributions are not corrected for acceptance or 
    efficiency effects.
    The vertical line is drawn at 135\,MeV/c$^2$.
    The absolute 
    gain calibration of the pEEMC has been determined by the position 
    of the mass peak in spin-summed distributions.
    Bin-by-bin constraints are applied to the energy sharing parameter
    to reduce background at small invariant mass.  
    The filled area represents events collected when the yellow spin 
    polarization direction is up, while unfilled represents spin down.  
    There is negligible dependence of the peak position  
    on either spin or energy.
    }
  \label{mass_vs_energy}
\end{figure}

Neutral $\pi$ mesons were reconstructed 
with the pEEMC from two cluster events in the SMD
according to the formula,
\begin{equation}
  M_{\gamma\gamma}=E_\pi \sqrt{1-z^2_\gamma}\,
         \sin (\frac{\phi_{\gamma\gamma}}{2})
   \approx E_{tot}\sqrt{1-z^2_\gamma}\, \frac{d_{\gamma\gamma}}{2\, z_{vtx}}.
\end{equation}
The energy of the $\pi^0$, $E_\pi$, was taken to be the total 
energy deposited in all of the towers in the calorimeter, $E_{tot}$.
The opening angle between the photons, $\phi_{\gamma\gamma}$, was 
determined by the measurement of two values:  the vertex position, 
$z_{vtx}$, given by the time difference measured by the east and west 
STAR BBC's, and the distance between the two photons at the calorimeter, 
$d_{\gamma\gamma}$.
Both $d_{\gamma\gamma}$ and the di-photon energy sharing parameter,
$z_{\gamma}=|E_1-E_2|/(E_1+E_2)$,
were measured by an analysis of the energy deposited in the strips 
of the two orthogonal SMD planes.
Typical events had these SMD distributions fit with two peaks used to
model the transverse profile of the electromagnetic shower from the
incident photons.
The value of $d_{\gamma\gamma}$ was determined from the fitted
centroids of the peaks, while $z_\gamma$ was determined from the 
fitted area under each peak.
Background at low invariant mass was reduced by constraining $z_\gamma$
as indicated in Figure~\ref{mass_vs_energy}, 
to ensure that both photons deposit significant energy in the SMD.
This algorithm resulted in a mass resolution of $20\,$MeV/c$^2$ for $\pi^0$ 
energies from $20-80\,$GeV, limited by the 
measurement of $\phi_{\gamma\gamma}$ (Fig.~\ref{mass_vs_energy}).
Due to the finite size of the collision diamond, making an assumption of
a fixed value for $z_{vtx,fixed}$ would result in a mismeasurement of 
$\phi_{\gamma\gamma}$.
The peak in the invariant mass distribution, reconstructed using 
$z_{vtx,fixed}$, was found to be linearly correlated with $z_{vtx}$ 
as determined from charged tracks 
reconstructed with the STAR time projection chamber, with 
which a subset of the FPD data was accumulated~\cite{akio}.
This provides evidence that the observed $\pi^0$ mesons were produced 
in $p+p$ collisions.

The absolute energy scale for each tower was determined from the $\pi^0$ 
peak in the invariant mass distribution.
The invariant mass was sorted according to the calorimeter tower with the 
greatest energy deposition in each event, and then the gain for each 
tower was adjusted to match the known mass of the $\pi^0$ meson.
Since typical events involve multiple calorimeter towers, the gain
matching procedure was iterative.
After approximately five iterations, this procedure converged to a 
stable set of values for each fill with an absolute uncertainty better 
than 1\%.
Small drifts of the gain on the order of a few percent were observed for 
many towers.
It has been checked that the position of the $\pi^0$ peak had negligible
dependence on the spin-state of the yellow beam and was independent
of $\pi^0$ energy, as shown in Figure~\ref{mass_vs_energy}.
The energy calibration of the Pb-glass arrays were also performed
with $\pi^0$ mesons, although the mass resolution was significantly 
worse since the positions of the photons at the detector were not as 
well measured.

\begin{figure}
  \includegraphics[height=2.5in]{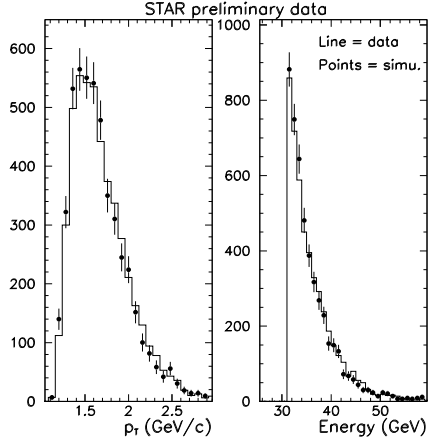}
  \caption{The spin-summed transverse momentum and energy distributions
           for events seen with the Pb-scintillator sampling 
           calorimeter (pEEMC), uncorrected for efficiency or acceptance
           effects.
           The histogram is data from a single fill subjected to the 
           data analysis algorithm described in the text.
           The points are a Monte-Carlo simulation using events generated 
           from a PYTHIA minimum-bias sample together with a GEANT model
           of the pEEMC, subjected to identical analysis constraints as
           the data.
           The simulation agrees well with the data for virtually all
           observables, indicating minimal contributions from background
           sources other than $p+p$ collisions.
    }
  \label{data_sim}
\end{figure}

The FPD data were compared with a simulation of $p+p$ collisions 
using PYTHIA together with a full GEANT simulation of 
the pEEMC response.
Minimum-bias PYTHIA events with more than $25\,$GeV of energy within a 
box of size comparable to the pEEMC were run through GEANT.
The simulated detector responses were processed through the analysis 
algorithm as if they were data.
The simulation was found to compare well with the data for an 
over-determined set of kinematic variables, bolstering the evidence that
the FPD was measuring $\pi^0$ mesons resulting from $p-p$ collisions.
A comparison of the data and the simulation for the $p_T$ and the
energy spectra can be seen in Figure~\ref{data_sim}.

The orientation of the yellow beam polarization for FPD triggered events 
was 
determined by measuring the time difference between the FPD trigger and
spin direction bits provided by RHIC.
The relative luminosity of collisions with polarization direction of
the yellow beam oriented up or down was measured by counting 
the coincidences of charged particles fore and aft of the 
collision vertex by the STAR BBC sorted by the yellow beam spin
direction bits~\cite{joanna}.
The live time of the FPD data acquisition system was 
measured by counting the number of events acquired divided
by the number of events which satisified the trigger condition.
No appreciable spin-dependence of the live time was observed.
Correcting the $\pi^0$ yield by the luminosity and live time
resulted in a normalized $\pi^0$ yield with fill-to-fill 
stability on the order of 15\%.
The value of the relative luminosity correction ($R$ in 
Equation~\ref{asym}) was typically on the order of $1.15$, and is 
understood to come from variations in the beam intensity from
bunch-to-bunch~\cite{joanna}.

\section{Results}

The measured analyzing power is not strongly affected by cuts used to 
identify $\pi^0$ mesons. 
The $A_N$ for the $\pi^0$ candidate events 
in the mass range $70<M_{\gamma\gamma}<300\,$MeV/c$^2$ shown in 
Figure~\ref{mass_vs_energy} is consistent with $A_N$ for 
the energy spectra observed with the pEEMC.
The value of Feynman-$x$ is approximately $E_{tot}/100\,$GeV.
A preliminary analysis of the simulation in Figure~\ref{data_sim} indicates
that events which trigger the FPD are composed of 95\% photons,
95\% of which are daughters from $\pi^0$ decay.
Non-photon triggers predominantly come from hadron showers, while other 
photon triggers mostly come from other meson decays, such as the 
$\omega,\ \eta$, and $\eta^\prime$.

\begin{figure}
  \includegraphics[height=3.5in]{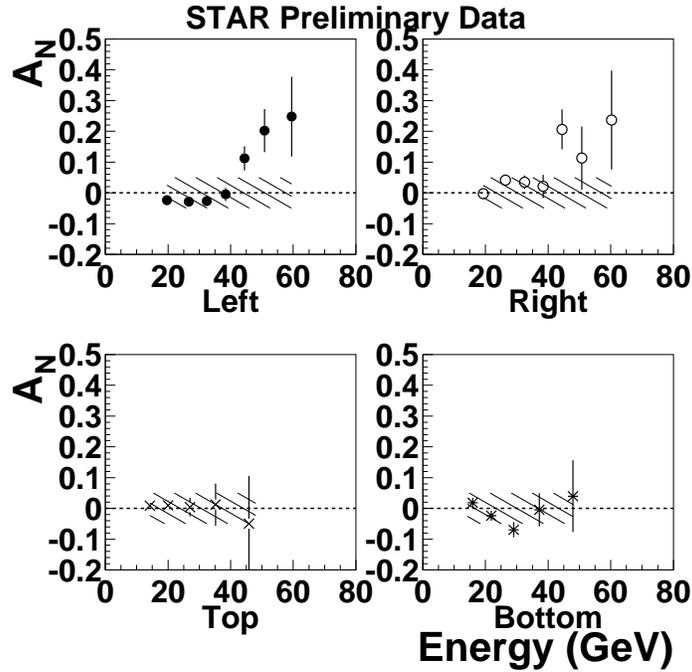}
  \caption{Preliminary results of the analyzing power for the 
           energy spectra in the four arms of the FPD detector 
           measured with
           a vertically polarized proton beam.
           The lines on the data points represent the statistical 
           uncertainty, while the hatched area represents an estimate
           of the systematic uncertainty.
           The results from the Pb-scintillator sampling calorimeter to
           beam-left are consistent with the Pb-glass
           array to beam-right, while the Pb-glass arrays above and below
           the beam are consistent with zero.
           The analyzing power for identified $\pi^0$ mesons with
           the Pb-scintillator sampling calorimeter is consistent
           with these data, but with significantly larger statistical
           uncertainties.
           The size and shape of the analyzing power is similar to that
           seen for $\pi^0$ mesons at E704~\cite{E7041,E7042}.
    }
  \label{allDetectors}
\end{figure}

The analyzing power for the energy spectra in the pEEMC is shown in
the upper-left plot in Figure~\ref{allDetectors}.
Also displayed in Figure~\ref{allDetectors} is $A_N$ for the energy 
spectra measured with the Pb-glass arrays.
The analyzing power observed on the left side with the Pb-scintillator 
sampling calorimeter is consistent with $A_N$ measured with the Pb-glass 
array on the right side, even though systematic effects arising from 
hadronic contributions in these two detector arms are different.
The analyzing powers observed above and below the beam pipe with Pb-glass
arrays are consistent with zero.

The preliminary systematic uncertainty is taken to be constant
throughout the energy range covered by the detectors, its value being
$\delta A_N = 0.05$.
This estimate has three primary, approximately equal, components:
the average difference between the 
left and right detectors, the difference between the energy 
spectra asymmetry and the asymmetry for identified $\pi^0$ mesons, 
and the time dependence of the spin-dependent asymmetry 
seen with the pEEMC.
The systematic uncertainty does not include the asymmetric
normalization uncertainty from the beam polarization.

In summary, we have observed that the analyzing power for $\pi^0$ mesons at 
$\sqrt{s}=200\,$GeV is similar in magnitude and $x_F$ dependence to that
measured at collision energies an order of magnitude smaller.

\bibliographystyle{aipproc}   

\end{document}